\RequirePackage{fix-cm}
\documentclass[smallextended]{svjour3}       
\smartqed  
\usepackage{graphicx}
\usepackage{graphicx}
\usepackage{amsmath}
\usepackage{amsfonts}
\usepackage{mathrsfs}
\usepackage{amssymb}
\usepackage{graphicx}
\usepackage{epsfig}
\usepackage{hyperref}
\usepackage{amscd}
\usepackage{color}
\newcommand{\be}{\begin{equation}}
\newcommand{\ee}{\end{equation}}
\newcommand{\bea}{\begin{eqnarray}}
\newcommand{\eea}{\end{eqnarray}}

\begin{document}

\title{Comparison of Vacuum Static Quadrupolar  Metrics
}


\author{Francisco Frutos-Alfaro \and Hernando Quevedo \and Pedro S\'anchez
}


\institute{Francisco Frutos-Alfaro \at
	School of Physics and Space Research Center of the University of Costa Rica
	\and
	Hernando Quevedo \and Pedro S\'anchez \at
	Instituto de Ciencias Nucleares, Universidad Nacional Aut\'onoma de 
M\'exico,  Ciudad de M\'exico 04510, Mexico
	\and
	Hernando Quevedo \at
	Dipartimento di Fisica and ICRA, Universit\`a di Roma ``La Sapienza",  I-00185 Roma, Italy
	\and
	Hernando Quevedo \at
	Department of Theoretical and Nuclear Physics, Kazakh National University, Almaty 050040, Kazakhstan
}


\date{Received: date / Accepted: date}

\maketitle

\begin{abstract}
We investigate the properties of static and axisymmetric vacuum solutions of 
Einstein equations which generalize the Schwarzschild spherically symmetric 
solution to include a quadrupole parameter. We test all the solutions with 
respect to elementary and asymptotic flatness and curvature regularity. 
Analyzing their multipole structure, according to the relativistic invariant 
Geroch definition, 
we show that all of them are equivalent up to the level of the quadrupole. 
We conclude that the $q-$metric, a variant of the Zipoy-Voorhees metric, is
the simplest generalization of the Schwarzschild metric, containing a 
quadrupole parameter. 
\end{abstract}

\section{Introduction}

Most applications of Einstein's gravity theory follow from the
investigation of exact solutions of the corresponding field equations. 
In the case of relativistic astrophysics, asymptotically flat solutions in 
empty space are of particular importance in order to 
describe the physical properties of the exterior field of compact objects 
\cite{solutions}. From a physical point of view,  
it is sufficient in this case to limit ourselves to static and stationary 
solutions which are axially symmetric. 
In addition, it is appropriate to classify them in accordance with certain 
criteria which permits a comparison of their main properties. 
Using the analogy with Newtonian gravity, we propose to classify them in 
terms of their multipole moments.

The problem of defining invariant multipole moments in general relativity 
was first solved by Geroch and Hansen (GH) \cite{Geroch,Hansen}, 
who proposed definitions for mass and spin multipoles of asymptotically flat 
spacetimes in vacuum. Moreover, Thorne, Simon and Beig defined relativistic 
multipole moments \cite{Simon,Thorne} for non-stationary spacetimes. 
A proof of the equivalence between the GH moments and the Thorne moments for 
stationary systems was provided by G\"ursel \cite{Guersel}. An elegant 
method to derive explicit expressions for the multipole moments of a given 
stationary and axially symmetric spacetime with asymptotic flatness was 
found by Fodor, Hoenselaers and Perj\'es (FHP) \cite{Fodor} using the Ernst 
formalism. This FHP method was generalized by Hoenselaers and Perj\'es 
\cite{Hoenselaers}. Finally, Ryan found an alternative method for deriving 
the relativistic multipole moments \cite{Ryan} which has been intensively 
applied in relativistic astrophysics.

Although for the study of the gravitational field of relativistic compact 
objects, it is necessary to consider stationary solutions that take into 
account the rotation of the source, in this work, we will focus on the 
study of the static case to explore in detail the physical properties of 
the solutions which then will be  generalized to the case of stationary 
fields. From a physical point of view, the most important multipoles of a 
mass distribution are the monopole and the quadrupole; in this work, we 
will focus our analysis on mainly these two multipoles.

The first solution with only monopole moment was derived by Schwarz\-schild in 
1916, just a couple of months after the publication of the theory of general 
relativity \cite{schw16}.  In 1917, Weyl found a class of static and 
axisymmetric solutions to the vacuum Einstein field equations 
\cite{Weyl,Weyl2}. The first static solution with quadrupole moment which 
includes the Schwarzschild metric as special case was found by Erez and 
Rosen in 1959 \cite{DZN,ER}. This quadrupolar solution was generalized to 
include an infinite mumber of multipole moments by Quevedo in 1989 
\cite{Quevedo1}. In 1966 and 1970, Zipoy and Voorhees found a transformation which allows us to generate
new static solutions from known solutions \cite{Voorhees,Zipoy}. In particular, applying this transformation 
to the Schwarzschild metric, one obtains a new solution which,  after a redefinition of the Zipoy-Voorhees parameter, 
was interpreted as the simplest static solution with generalizes the 
Schwarzschild metric and includes a quadrupole moment ($ q $-metric) \cite{Quevedo3}. 
In 1985, Gutsunaev and Manko found an exact solution with monopole and 
quadrupole moments which was shown in \cite{quev87} to have the same 
quadrupole as in the Erez-Rosen metric, but different contributions to 
higher relativistic multipole moments. In 1990, Manko \cite{man90} found a quadrupolar metric which can be interpreted as 
the non-linear combination of the Schwarzschild monopole solution with the quadrupolar term of the Weyl solution. 
In 1994, Hern\'andez-Pastora and 
Mart{\'{\i}}n \cite{HM} derived two  exact solutions with different 
monopole-quadrupole structures.

To our knowledge, the above list includes all known static and 
asymptotically flat solutions of Einstein's equations in empty space. 
The main goal of the present work is to investigate the most important physical 
properties of these solutions. In particular, we will analyze the elementary 
flatness condition, curvature singularities, multipole moments structure and the 
relationships between them.

This paper is organized as follows.  In Sec. \ref{sec:gen}, we present the 
general line element for static axisymmetric spacetimes and the 
corresponding vacuum field equations, and review the most general 
aymptotically flat solution in cylindrical coordinates discovered by Weyl.
In Sec. \ref{sec:static}, we present 
the solutions that contain the Schwarzschild spacetime as a particular case 
and an additional parameter which determines the quadrupole of the 
gravitational source. Then, in Sec. \ref{sec:cond}, we investigate the 
conditions that the solutions must satisfy in order to be able to describe 
the exterior gravitational field of compact objects. Sec. \ref{sec:mms} is 
devoted to the study of the multipole structure of the solutions. Finally, 
in Sec. \ref{sec:con}, we discuss our results and present some initiatives 
for future works.


\section{General properties of static and axisymmetric vacuum solutions}
\label{sec:gen}

Although there exist in the literature many suitable coordinate systems, 
static axisymmetric gravitational fields are usually described
in cylindrical coordinates $(t,\rho,z,\varphi)$, following the seminal work of 
Weyl. Stationarity implies that there exists a timelike Killing vector field 
with components
$\delta^\alpha_t$, i.e.,  $ t $ can be chosen as the time coordinate and 
the metric does not depend on time, 
$ \partial g_{\alpha\beta}/\partial t = 0 $. 
Axial symmetry, in addition, implies
the existence of a spacelike Killing vector field with components 
$ \delta^\alpha_\varphi $,
which commutes with the timelike Killing vector. The coordinates can then be 
chosen such 
that $ \partial g_{\alpha\beta}/\partial \varphi = 0 $, and the axis of 
symmetry 
corresponds to 
$\rho=0$. Furthermore, if we assume that
the timelike Killing vector is hypersurface-orthogonal, the spacetime is 
static, i.e.,
it is invariant with respect to the transformation 
$ \varphi \rightarrow -\varphi $.

Furthermore, using  the properties of staticity and axial symmetry, together
with the vacuum field equations, for a general metric of the form 
$ g_{\alpha\beta} = g_{\alpha\beta}(\rho, \, z) $, it is possible to show that 
the most general line element for this type of gravitational fields can be 
written in the Weyl-Lewis-Papapetrou form as \cite{Weyl,lewis,pap,solutions}
\be
\label{lel1}
ds^2 = {\rm e}^{2\psi} dt^2 - {\rm e}^{-2\psi} 
\left[{\rm e}^{2 \gamma}(d\rho^2+dz^2) + \rho^2d \varphi^2\right] \ ,
\ee
where $\psi$ and $\gamma$ are functions of $\rho$ and $z$ only. The vacuum 
field equations can be reduced to the following set of independent 
differential equations  
\be
\label{eqpsi}
\psi_{\rho\rho}  + \frac{1}{\rho}  \psi_\rho + \psi_{zz} = 0 \ ,
\ee
\be
\label{eqkstatic}
\gamma_\rho =  \rho\left( \psi_\rho ^2 - \psi_z ^2\right)\ ,
\quad 
\gamma_z  = 2 \rho  \psi_\rho \psi_z  \ ,
\ee
where $\psi_\rho = \partial \psi/\partial \rho$, etc. We see that the main 
field equation (\ref{eqpsi}) corresponds to the linear Laplace equation for 
the metric function $ \psi $. Furthermore, the solution for the function 
$ \gamma $ can be obtained by quadratures once the function $\psi$ is known.

The general solution of Laplace's equation is known and, if we demand 
additionally asymptotic flatness, we obtain the Weyl solution 
\cite{Weyl,solutions} 
\be
\label{weylsol}
\psi = \sum_{n=0}^\infty \frac{a_n}{(\rho^2+z^2)^\frac{n+1}{2}} 
P_n({\cos\theta}) \ ,
\qquad \cos\theta = \frac{z}{\sqrt{\rho^2+z^2}} \ ,
\ee
where $ a_n $ $ (n = 0 , \, 1 , \, ...) $ are arbitrary real constants, and 
$ P_n(\cos{\theta})$ represents the Legendre polynomials of degree $ n $. 
The expression for the metric function $ \gamma $ can be obtained from the 
two first-order differential equations (\ref{eqkstatic}). Then 
\be
\gamma = - \sum_{n,m=0}^\infty \frac{a_na_m (n+1)(m+1)}{(n+m+2)
(\rho^2+z^2)^\frac{n+m+2}{2}}
\left(P_n P_m - P_{n+1} P_{m+1} \right) \ .
\ee
Since this is the most general static, axisymmetric, asymptotically flat
vacuum solution, it must contain all known solutions of this class. 
In particular, one of the most interesting special solutions, which is 
Schwarzschild's spherically symmetric black hole spacetime, must be included 
as a special case. To see this, we must choose the constants $ a_n $ in such 
a way that the infinite sum (\ref{weylsol}) converges to the Schwarzschild 
solution in cylindrical coordinates. A straightforward computation shows that
\be
a_{2n} = - \frac{m^{2n+1}}{2n+1}, \quad a_{2n+1} = 0\ ,
\ee
where $m$ is the mass parameter \cite{TJLHP}. Clearly, this representation is 
not appropriate to handle the Schwarzschild  metric.

It turns out that  to investigate the properties of solutions with multipole 
moments, it is convenient to use prolate spheroidal coordinates 
$ (t, \, x, \, y, \, \varphi) $ in which the line element can be written as
\bea
\label{lelxy}
ds^2 & = & {\rm e}^{2 \psi} dt^2 \nonumber \\
& - & {\sigma^2}{\rm e}^{- 2 \psi} 
\left[ {\rm e}^{2 \gamma} (x^2 - y^2)\left(\frac{dx^2}{x^2 - 1} 
+ \frac{dy^2}{1 - y^2} \right) + (x^2 - 1)(1 - y^2) d\varphi^2\right] , 
\eea
where
\be 
x = \frac{r_+ + r_-}{2 \sigma} \ ,\quad (x^2 \geq 1), \quad 
y = \frac{r_+-r_-}{2 \sigma} \ , \quad (y^2 \leq 1) 
\ee
\be 
r_\pm^2 = \rho^2 + (z\pm \sigma)^2 \ , \quad \sigma = {\rm const.} \ ,
\ee 
and the metric functions $ \psi $, and $ \gamma $ depend on $ x $ and $ y $, 
only. In this coordinate system, the field equations become
\be
\label{eqpsixy}
[(x^2 - 1) \psi_x]_x + [(1 - y^2) \psi_y]_y = 0 \ ,
\ee
\bea
\label{eqgamxy}
\gamma_x & = & \left(\frac{1 - y^2}{x^2 - y^2} \right)
\left[ x (x^2 - 1) 
\psi_x^2 
- x (1 - y^2) \psi_y^2 
- 2 y (x^2 - 1) \psi_x 
\psi_y \right] , \\
\gamma_y & = & \left(\frac{x^2 - 1}{x^2 - y^2}\right)
\left[ y (x^2 - 1) 
\psi_x^2 
- y (1 - y^2) \psi_y^2 
+ 2 x (1 - y^2) \psi_x 
\psi_y \right] . \nonumber
\eea
The simplest physically meaningful solution to the above system of 
differential equations is the Schwarzschild solution
\be
\psi_{_S} = \frac{1}{2}\ln\left(\frac{x-1}{x+1}\right) \ ,\quad
\gamma_{_S} = \frac{1}{2}\ln\left(\frac{x^2-1}{x^2-y^2}\right)\ ,
\label{schw}
\ee
which takes the standard form in spherical coordinates with $x=r/m-1$, 
$y=\cos\theta$, and $\sigma=m$.  
In principle, there could be an infinite number of exact solutions to the 
above equations. Not all of them, however, can be physically meaningful, in particular, 
if we demand that they should describe the exterior field of realistic compact 
objects. To this end, it is necessary that the solutions satisfy the 
conditions of asymptotic flatness, elementary flatness and regularity. 

Asymptotic flatness means that at spatial infinity, the solution reduces to 
the Minkowski metric, indicating that the gravitational field far away from 
the source is practically negligible. This is a consequence of the long-range 
property of the gravitational interaction. In the case of the static metric in prolate 
spheroidal coordinates \ref{lelxy}, this condition implies that 
\be
\lim_{x\to\infty} \psi = {\rm const.}, \quad 
\lim_{x\to\infty} \gamma = {\rm const.} \ ,
\ee
where the constants can be set equal to zero by a suitable rescaling of the coordinates.

Elementary flatness is necessary in 
order to guarantee that near the rotation axis the geometry is Lorentzian, 
i.e., there are no conical singularities on the axis \cite{solutions}. 
This condition can be expressed in an invariant manner by using the 
spacelike  Killing vector field $\eta^\alpha = \delta^\alpha_\varphi$ as 
\be
\lim_{\rho \to 0} \frac{(\eta^\alpha\eta_\alpha)^{,\beta}
(\eta^\alpha\eta_\alpha)_{,\beta}}{ 4(\eta^\alpha\eta_\alpha)} = 1\ .
\label{elflat}
\ee 
A direct computation by using the general line element in prolate spheroidal coordinates shows that the elementary flatness 
condition is equivalent to demanding that
\be
\lim_{y\to\pm 1} \gamma = 0 \ ,
\ee
independently of the value of the spatial coordinate $x$.

Finally, the regularity condition implies that the solution must be free of 
curvature singularities outside a region located near the origin of coordinates so 
that it can be covered by an interior solution. Curvature singularities can be detected by 
analyzing the behavior of curvature invariants. In general, the Riemann curvature tensor
in four dimensions possess 14 independent invariants. In the case of vacuum spacetimes, however,
the Riemann tensor coincides with the Weyl tensor that has only four invariants which can be expressed
as \cite{cm91} 
\bea
K & = & I_1 = R_{\alpha\beta\gamma\delta} R^{\alpha\beta\gamma\delta} \ ,\quad
I_2 = * R_{\alpha\beta\gamma\delta}  R^{\alpha\beta\gamma\delta} \ ,\nonumber\\ 
I_ 3 & = & R_{\alpha\beta\gamma\delta}  R^{\gamma\delta\lambda\tau} R_{\lambda\tau}^{\ \ \alpha\beta} \ , \quad
I_4 = * R_{\alpha\beta\gamma\delta}  R^{\gamma\delta\lambda\tau} R_{\lambda\tau}^{\ \ \alpha\beta} \ ,
\eea
where the dual is defined as 
\be
* R_{\alpha\beta\gamma\delta} = \frac{1}{2} \epsilon_{\alpha\beta\lambda\tau} R^{\lambda\tau}_{\ \ \gamma\delta} \ , 
\ee
with $\epsilon_{\alpha\beta\lambda\tau}$ being the Levi-Civita symbol. The quadratic invariants $K=I_1$ and $I_2$ are usually known as the
Kretschmann and the Chern-Pontryagin scalars, respectively. If anyone of the four invariants happens to diverge at some particular place, 
it is said that there exists a curvature singularity at that place. 

In the next section, we will investigate the properties of several exact 
solutions with monopole and quadrupole moment. In particular, we will find out if they satisfy 
all the conditions to be physically relevant in the sense that they can be used to describe the exterior gravitational field of
compact objects.


\section{Static Vacuum Metrics with Quadrupole}
\label{sec:static}

As mentioned in the last section, the Weyl metric can be considered as the 
most general static and axisymmetric solution which contains an infinite 
number of parameters, representing all the multipole moments. Therefore, a 
particular choice of parameters could represent a solution with only mass 
and quadrupole. However, such a form of a metric with an infinite number of 
parameters is not very suitable to be applied in the case of realistic 
sources like compact astrophysical objects. For this reason, we consider now 
metrics which include only two independent parameters that can be 
interpreted as mass and quadrupole. 

In 1959, Erez and Rosen \cite{ER} presented a solution which generalizes 
the Schwarzschild metric and contains an additional parameter $ q $. 
In this case, the function $\psi$ can be expressed as
\be
\psi_{_{ER}} = \frac{1}{2}\ln\left(\frac{x-1}{x+1}\right) + \frac{1}{2} 
q(3y^2-1)\left[ \frac{1}{4}(3x^2 -1) \ln\left(\frac{x-1}{x+1}\right)
+\frac{3}{2}x\right]\ .
\ee
The corresponding function $\gamma_{_{ER}}$ cannot be expressed in a compact form 
and is given explicitly in Appendix \ref{sec:gamma}. This 
solution was obtained by using the method of separation
of variables for the function $\psi$. An explicit generalization which 
contains higher multipole moments was presented in 1989 in 
\cite{Quevedo1} by using the same method.

In 1984, Gutsunayev and Manko \cite{GM} found a new static solution for the 
function $\psi$ which is given by
\be
\psi_{_{GM}} = \frac{1}{2}\ln\left(\frac{x-1}{x+1}\right) 
+ {q} \frac{x}{(x^2 - y^2)^3} (x^2 - 3 x^2 y^2 + 3 y^2 - y^4) \ ,
\ee
and the function $\gamma_{GM}$ is given explicitly in  Appendix \ref{sec:gamma}. 
This 
solution was found by applying a particular differential operator 
to the Schwarzschild metric. This method was shown to be based upon the 
property that in Cartesian coordinates the derivatives of a harmonic 
function are also harmonic functions \cite{quev87}.  

In 1990, Manko \cite{man90} found a different  static solution in the form
\be
\psi_{_M} = \frac{1}{2}\ln\left(\frac{x-1}{x+1}\right) 
 + q \frac{ 3x^2 y^2 -x^2 -y^2+1}{2(x^2+y^2-1)^{5/2}}\ ,
\ee
which leads to a particular function $\gamma_M$ given in Appendix \ref{sec:gamma}.
The first term of this solution corresponds to the Schwarzschild metric, whereas the second term coincides 
with the quadrupolar term of the general Weyl solution in prolate spheroidal coordinates.

Furthermore, in 1994, Hern\'andez-Pastora and Mart{\'{\i}}n \cite{HM} derived 
two different exact solutions which can be written as
\bea
\label{psihm1}
\psi_{_{HM1}} & = &\frac{1}{2}\ln\left(\frac{x-1}{x+1}\right) 
- \frac{5}{8} {q} \left[\frac{1}{4} \left((3 x^2 - 1) (3 y^2 - 1) - 4 \right) 
\ln\left(\frac{x-1}{x+1}\right) \right. \nonumber \\
& + & \left. \frac{2 x}{(x^2 - y^2)} - \frac{3}{2} x (3 y^2 - 1) \right] 
\eea
and 
\bea
\label{psihm2}
\psi_{_{HM2}}  & = & \psi_{_{HM1}}
- \frac{5}{32} {q}^2 \left[\left({33} + {90} P_2(x) P_2(y) 
- \frac{153}{2} P_4(x) P_4 (y) \right) \ln\left(\frac{x-1}{x+1}\right) 
 \right. \nonumber \\
& - & \left. {135} x P_2(y) 
- \frac{153}{24} x (55 - 105 x^2) P_4(y) \right. \\
& - & \left. \frac{x}{x^2 - y^2}  
\left(33 - \frac{5}{(x^2 - y^2)^2} (3 x^2 y^2 + y^4 - x^2 - 3 y^2) \right) 
\right] , \nonumber
\eea
where 
\bea
P_2(y) & = & \frac{1}{2} (3 y^2 - 1) , \\
P_4(y) & = & \frac{1}{8} (35 y^4 - 30 y^2 + 3) . 
\eea
Here we have corrected several typos which are present in the original publications. 
The corresponding functions $\gamma_{_{HM1}}$ and $\gamma_{_{HM2}}$ have a quite complicated 
structure which we present explicitly in Appendix \ref{sec:gamma}. 

Finally, in 1966 and 1970 Zipoy \cite{Zipoy} and Voorhees \cite{Voorhees}, 
respectively, found a particular symmetry of the vacuum field equations, and derived a 
transformation which can be used to generate new solutions from known solutions. In the case of the Schwarzschild metric,
the new solution can be expressed simply as  
\be
\psi_{_{ZV}} = \frac{1}{2}\delta \ln\left(\frac{x-1}{x+1}\right) \ , \quad
\gamma_{_{ZV}} = \frac{1}{2}\delta^2 \ln\left(\frac{x^2-1}{x^2-y^2}\right) \ ,
\ee
where $\delta$ is an arbitrary real constant. This solution is also known 
as the $\delta-$metric of the $\gamma-$metric for notational reasons \cite{mala04}. Later 
on, in 2011, this metric was reinterpreted as a quadrupolar metric and 
renamed as the $q-$metric \cite{Quevedo3} which in spherical coordinates 
can be transformed into the simple form
\begin{eqnarray}
\label{ZV}
d {s}^2 & = & \left(1 - \frac{2 m}{r}\right)^{1+q} d {t}^2  
- \left(1 - \frac{2 m}{r}\right)^{-q} \\
& \times & \left[\left(1 + \frac{m^2\sin^2\theta}{r^2 - 2 m r} 
\right)^{-q(2+q)} 
\left(\frac{d {r}^2}{1 - {2 m} / {r}} + r^2 d {\theta}^2\right) 
+ r^2 \sin^2{\theta} d {\varphi}^2 \right] . \nonumber
\end{eqnarray}

It is easy to see that all the above solutions represent a generalization 
of the Schwarzschild metric which is obtained in the limiting case 
$q\to 0$. To our knowledge, the solutions presented above are the only exact 
solutions that generalize the Schwarzschild monopole solution and satisfy 
the conditions expected from a metric that describes a realistic 
gravitational field.


\section{Physical conditions}
\label{sec:cond}

All the solutions presented in the last section are asymptotically flat 
because at spatial infinity they behave as 
\be
\psi_{_0}=  \lim_{x\to\infty} \psi = 0 \ , \quad
\gamma_{_0}=  \lim_{x\to\infty} \gamma = 0 \ , 
\ee
which determine the Minkowski metric, independently of the value of $y$. Notice, moreover, that this condition is satisfied for all finite values of the independent parameters $m$ and $q$. This means that for any finite values of the monopole and quadrupole moments, the solutions presented in the last section are asymptotically Minkowski. 

As mentioned above, the condition that no conical singularities exist on the symmetry 
axis (\ref{elflat}) in prolate spheroidal coordinates 
becomes
\be
\lim_{y\to \pm 1} \gamma = 0 \ .
\ee
An inspection of the $\gamma$ function for the Erez-Rosen, Gutsunayev-Manko and Manko solutions and the $q-$metric, mentioned in the 
last section, shows that this condition is always satisfied, independently of 
the value of $x$, indicating that all of them are elementary flat. In the case of the Hern\'andez-Mart\'\i n solutions, however, a 
direct computation shows that they are elementary flat only for positive values of the coordinate $x$. In spherical coordinates, this means that the HM solutions are well-defined only outside the radius $r=2m$. A geometric and physical  analysis inside the horizon $r=2m$ is possible only by considering the presence of conical singularities along the symmetry axis.

We now analyze the regularity condition by using first the Kretschmann scalar 
$K= R_{\alpha\beta\gamma\delta} R^{\alpha\beta\gamma\delta}$.
Fist, we consider the Schwarzschild metric (\ref{schw}) for which we obtain
\be
K_{_S} =  \frac{48}{m^4 (x+1)^6}.
\label{kschw}
\ee 
This expression is singular only for $x=-1$ ($r=0$), indicating the 
well-known fact that the Schwarzschild spacetime is singular only at 
the origin of coordinates.

Another example of a solution that can be investigated analytically is the $q-$metric. 
In this case, all the calculations can be 
performed explicitly and the resulting Kretschmann scalar reads
\begin{equation}
K_q  = \frac{48}{\sigma^4} \, (q+1)^2 \frac{p(x,y;q)}{(x+1)^{2(q^2+3q+3)} 
(x-1)^{2(q^2+q+1)} (x^2-y^2)^{-2q^2-4q+1}} \ ,
\end{equation}
where  
\bea
p(x,y;q) & = & (x-1)^2 (x^2-y^2) - 2q(x-1)^2(x+y^2) \\
& + & q^2\left[ (2-y^2)x^2-3(1-y^2)x+\frac{1}{2}(4-7y^2) \right] \nonumber \\
& - & q^3\left( x-\frac{4}{3} \right)(1-y^2) + \frac{1}{3} q^4(1-y^2) . \nonumber
\eea
First, we see that for all values of $q$ there is always a singularity at 
$x=-1$. Moreover, we have two possible divergences at $x=1$ and 
$x=\pm y$. These divergent factors can only be canceled by the function $p$, 
but it does not vanish for 
$x=1$ or $x=\pm y$ for arbitrary values of $q$, except for $q=-2$. In this 
case, one has $p(x,y;-2) = (x+1)^2 (x^2-y^2)$ so that
\begin{equation}
K_{q=-2} = \frac{48}{\sigma^4 (x-1)^6} \ ,
\end{equation}
which diverges for $x=1$.  For other values of the parameter $q$, the 
Kretschmann scalar of the $q$-metric  diverges at $x=\pm 1$ 
and $x=\pm y$, as far as the exponents of the corresponding factors are 
negative. The exponents of the factors $x+1$ and $x-1$ are negative 
definite, but the exponent  of the factor $x^2-y^2$ vanishes for 
$q=-1+\sqrt{3/2}$ and $q=-1-\sqrt{3/2}$.

Consequently, the Kretschmann scalar of the $q$-metric diverges at $x=-1$ for 
$q \neq -2$, at $x=1$ for $q \neq 0$ and at $x = \pm y$ 
for $q \in (-1-\sqrt{3/2}, -1+\sqrt{3/2})$ restricted to $q \neq 0$ and 
$q \neq -2$. An additional restriction to the value of the parameter $q$ is 
imposed by assuming $\sigma > 0$ and requiring its mass monopole to be 
positive. We will see in the next section that this physical 
condition implies that $q>-1$, leading to the conclusion that the singularity 
at $x=-1$ is always present.

The investigation of the remaining quadrupolar solutions is much more complicated.
In Appendix \ref{sec:singer}, we present as an example the explicit analysis of the Erez-Rosen metric. 
The results of our analysis are summarized in Table \ref{table1}, where we include the Schwarzschild solution for
comparison, and use 
spherical coordinates with $x=r/m-1$ and $y=\cos\theta$.
The boldfaced radii represent singularities that are present, independently of 
the value of the parameters $m$, $q$ and the coordinate $\theta$. 
The remaining radii represent singularities 
which are not always present, but depend on the value of $q$ or the coordinate $\theta$. 
We see that only 
the $q-$metric is characterized by a completely singular horizon at $r=2m$, representing 
the outermost singularity, which is the only one that can be observed by an 
exterior observer. 
In the remaining cases, the Schwarzschild horizon remains partially regular, implying that for certain 
values of $q$, it is possible to observe the singularity located at  the origin of coordinates.  

Finally, we mention that the analysis of the remaining three curvature invariants does not lead to 
additional singularities.

\begin{table}
\begin{center}
\begin{tabular}{|c|c|}
\hline
Static metric          & Naked singularites  \\
\hline
Schwarzschild & $ \boldsymbol{ r = 0} $ \\
$q-$metric               & $ \boldsymbol{ r = 0}, \ m (1 \pm \cos{\theta}), \  \boldsymbol{2 m} $ \\
Erez-Rosen    & $ \boldsymbol{ r = 0}, \ m (1 \pm \cos{\theta}), \  2 m    $ \\
Gutsunayev-Manko & $ \boldsymbol{ r = 0}, \ m (1 \pm \cos{\theta}), \  2 m $ \\
Manko & $ \boldsymbol{ r = 0}, \ m (1 \pm \cos{\theta}), \  2 m $ \\
Hern\'andez-Mart\1n 1 and  2 \ \  & $ \boldsymbol{ r = 0}, \ m (1 \pm \cos{\theta}), \  2 m     $    \\
\hline
\end{tabular}
\end{center} 
\caption{Singularities of spacetimes with monopole and quadrupole moments. Boldfaced values are naked singularities which exist for all values of the parameters $m$, $q$ and $\theta$. Other singularities exist only for particular values of these parameters.  }
\label{table1}
\end{table}


\section{Multipole Moments}
\label{sec:mms}

Using the original definition formulated by Geroch \cite{Geroch}, the 
calculation of multipole moments is quite laborious. Fodor,
Hoenselaers and Perj\'es \cite{Fodor} found a relation between the Ernst 
potential \cite{Ernst1,Ernst2} and the multipole moments which facilitates 
the computation. In the case of static axisymmetric spacetimes, the Ernst potential is defined as
\be
\xi(x,y) = \frac{1- e^{2\psi}}{1+e^{2\psi}} \ .
\ee  
The idea is that the multipole moments can be obtained explicitly from the values of the Ernst potential on the axis by using
the following procedure.
On the axis of symmetry $y=1$, we can introduce the inverse of the Weyl 
coordinate $z$ as 
\be
\tilde z= \frac{1}{z} = \frac{1}{mx} \ , \quad {\rm with} \quad \sigma = m\ .
\ee
If we now introduce the inverse potential as 
\be
\tilde \xi (\tilde z, 1) = \frac{1}{\tilde z} \xi(\tilde z, 1) \ ,
\ee
the multipole moments can be calculated as 
\be
{\cal M}_n = m_n + d_n \ , \quad m _n = \frac{1}{n!} 
\frac{ d^n \tilde \xi (\tilde z, 1)}{d \tilde z ^n } \Big|_{\tilde z = 0} \ ,
\ee
where the additional terms $d_n$ must be determined from the original Geroch 
definition. The main point now is that the first term $m_n$ is completely determined by the $n-$th derivative
of the  inverse Ernst potential $\tilde\xi$, whereas the second term $d_n$ depends on the derivatives of order less than $n$, so 
that the moment ${\cal M}_n$ can be calculated explicitly once all the derivatives of order $n$ or less are known.
In  Appendix \ref{sec:mm}, we include the explicit expressions 
for the first ten additional terms.

In this manner, it is easy to show that for the Schwarzschild spacetime the multipole moments are given as 
\be
{\cal M}_0 = m \ , \quad {\cal M}_k = 0\ , \ (k\geq 1)\ ,
\ee 
a result which is in accordance with the  physical interpretation of the 
Schwarz\-schild metric obtained by using other methods.

For the  Erez-Rosen metric, we obtain 
\begin{eqnarray*}
\label{ERRMM}
{\cal M}_0 & = & m \\
{\cal M}_2 & = & Q \\
{\cal M}_4 & = & - \frac{2}{7} Q m^2 \\
{\cal M}_6 & = & - \frac{8}{231} Q m^4 \left(1 + 3 q \right) \\
{\cal M}_8 & = & - \frac{8}{3003} Q m^6 
\left(2 - \frac{74}{15} q + \frac{84}{45} q^2 \right) \\
{\cal M}_{10} & = & \frac{32}{3927} Q m^{8} \left(- \frac{28}{247}
+ \frac{37}{57} q + \frac{1124}{741} q^2 \right) ,
\end{eqnarray*}
where $ Q = {2} q  m^3 / {15} $. 

For the Gutsunayev-Manko metric, we obtain
\begin{eqnarray*}
\label{GMRMM}
{\cal M}_0 & = & m \\
{\cal M}_2 & = & Q \\
{\cal M}_4 & = & \frac{6}{7} Q m^2 \\
{\cal M}_6 & = & \frac{8}{231} Q m^4 \left(14 - 45 q \right) \\
{\cal M}_8 & = & \frac{8}{3003} Q m^6 \left(84 - 1282 q - 420 q^2 \right) \\
{\cal M}_{10} & = & \frac{32}{3927} Q m^8 \left(\frac{2772}{247} 
- \frac{1343804}{2717} q - \frac{50}{1463} q^2 \right) ,
\end{eqnarray*}
where $ Q = 2 q  m^3 $.

For the Manko solution, we obtain 
\begin{eqnarray*}
\label{MRMM}
{\cal M}_0 & = & m  \\
{\cal M}_2 & = & Q = - m^3 q \\
{\cal M}_4 &= & - \frac{8}{7} Q m^2 \\
{\cal M}_6 & = & \frac{1}{231} Q m^4 (180 q + 133)  \\
{\cal M}_8 & = & - \frac{2}{3003} Q m^6 (420 q^2 + 2182 q + 357) \\
{\cal M}_{10} & = & \frac{1}{969969} Q m^8 (1379100 q^2 + 1277710 q + 85701) .
\end{eqnarray*}

For the first Hern\'andez-Mart\'\i n metric, we obtain 
\begin{eqnarray*}
\label{HMRMMI}
{\cal M}_0 & = & m \\
{\cal M}_2 & = & Q \\
{\cal M}_4 & = & 0 \\
{\cal M}_6 & = & - \frac{60}{77} Q m^4 \\
{\cal M}_8 & = & - \frac{4}{3003} q Q m^6 \left(265 + 210 q \right) \\
{\cal M}_{10} & = & \frac{4}{3927} q Q m^8 
\left(- \frac{104370}{714} + \frac{769125}{1729} q \right) ,
\end{eqnarray*}
and for the second Hern\'andez-Mart\'\i n solution
\begin{eqnarray*}
\label{HMRMMII}
{\cal M}_0 & = & m \\
{\cal M}_2 & = & Q \\
{\cal M}_4 & = & 0 \\
{\cal M}_6 & = & 0 \\
{\cal M}_8 & = & - \frac{40}{143} q^2 Q m^6 \\
{\cal M}_{10} & = & - \frac{42140}{46189} q^2 Q m^8 ,
\end{eqnarray*}
where $ Q = q m^3 $.

Finally, for the $q-$metric we get 
\begin{eqnarray*}
\label{ZVRMM}
{\cal M}_0 & = &  \delta m \\
{\cal M}_2 & = & \frac{1}{3} \delta m^3 (1 - \delta^2) \\
{\cal M}_4 & = & \delta m^5
\left(\frac{19}{105} \delta^4 - \frac{8}{21} \delta^2 + \frac{1}{5} \right) \\
{\cal M}_6 & = & \delta m^7 \left(
- \frac{389}{3465} \delta^6 + \frac{23}{63} \delta^4 
- \frac{457}{1155} \delta^2 + \frac{1}{7} \right) \\
{\cal M}_8 & = & \delta  m^9 \left(
\frac{257}{3465} \delta^8 - \frac{44312}{135135} \delta^6 
+ \frac{73522}{135135} \delta^4 - \frac{54248}{135135} \delta^2 + \frac{1}{9} 
\right) \\
{\cal M}_{10} & = & \delta m^{11} \left(
- \frac{443699}{8729721} \delta^{10} + \frac{17389}{61047} \delta^8 
- \frac{27905594}{43648605} \delta^6 + \frac{6270226}{8729721} \delta^4 
\right. \\
& - & \left. \frac{5876077}{14549535} \delta^2 + \frac{1}{11}\right) \ ,
\end{eqnarray*}
where $\delta = 1+q$.

A comparison of these results show that all the above solutions are equivalent 
up to the quadrupole moment. Indeed, 
a simple redefinition of the parameter $q$ which enters all the metrics leads 
to equivalent values for the monopole and 
quadrupole moments. We see, however, that differences appear between all the 
solutions at the level of higher moments. The particularity of the first and second 
 Hern\'andez-Mart\'\i n solutions
is that by choosing the form of the metric $\psi_{_{HM}}$ appropriately,
the multipoles ${\cal M}_4$ and ${\cal M}_6$ can be made to vanish identically. 
This means that  by following the same procedure, it is possible to generate a solution 
with only monopole and quadrupole moments. In all the remaining solutions, contributions of
higher  multipoles are always present.

We conclude that from the point of view of the monopole-quadrupole structure 
all the  solutions presented in Sec. \ref{sec:static} are physically equivalent.



\section{Conclusions}
\label{sec:con}

In this work, we analyzed all the  exact solutions of Einstein's vacuum 
field equations which contain the Schwarzschild solution as a particular case 
and, in addition, possess an arbitrary parameter which determines the 
quadrupole of the gravitational source. In particular, we studied the 
Erez-Rosen, Gut\-suna\-yev-Manko, Manko, Hern\'andez-Pastora solutions and the 
$q-$metric, obtained from the Schwarzschild by applying a Zipoy-Voorhees transformation.

First, we established that all the above solutions are asymptotically and 
elementary flat. This means that at infinity the gravitational field strength 
is negligible, and the rotation axis is free of conical singularities, 
respectively. We performed also a detailed analysis of the Kretschmann scalar 
to determine the curvature singularity structure of these spacetimes. We found 
that in general there are three types of naked singularities which are located 
at the origin of coordinates $r=0$, between the origin and the Schwarzschild 
horizon 
$r= m(1\pm \cos\theta)$ and on the horizon $r=2m$, where $m$ is the mass of 
the gravitational source. The main difference is that only 
in the case of the $q-$metric, the outermost singularity located at $r=2m$ 
exists for all values of the parameters $m$ and  $q$ and the coordinate $\theta$.
  For all the remaining 
metrics, the second and third singularities exist only for certain specific 
values of $q$ or $\theta$. This means that in principle it is possible to observe the 
interior singularities located at $r=0$ and $r=m(1\pm \cos\theta)$, which is 
not possible in the case of a spacetime described by the $q-$metric. Suppose 
that we want to use an interior solution to  ``cover" the naked singularities 
generated by the quadrupole. In the case of the $q-$metric, the surface of the 
interior mass distribution can be located anywhere outside the outermost 
singularity situated at $r=2m$. In the case of all the remaining exterior 
metrics, the surface of the interior distribution can have even a zero radius 
for certain values of the quadrupole parameter. 

The study of the multipole moments of all the solutions shows that by choosing the 
quadrupole parameter appropriately all of them are characterized by the same 
mass and quadrupole, although differences can appear at the level of higher 
multipoles. This means that all the 
solutions can be used to describe the exterior gravitational field of a 
distorted mass distribution with quadrupole moment.

Our results show that all the solutions analyzed in this work are equivalent 
from the physical point of view in the sense that they satisfy
all the conditions that are necessary to describe the exterior gravitational 
field of realistic compact objects. Nevertheless, from a practical point of 
view the $q-$metric presents certain advantages over the remaining metrics. 
Indeed, the mathematical structure of this metric is very simple which 
facilitates its study. For instance, when searching for interior solutions 
with quadrupole that could be matched with an exterior quadrupolar metric, 
one certainly would try first the $q-$metric because of its simplicity. 

To completely describe the gravitational field of realistic compact objects 
with qua\-dru\-pole, it is necessary to take the rotation into account. Moreover, 
a suitable interior solution is also necessary in order to describe the entire 
spacetime, as required in general relativity. Due to the mathematical 
complexity of the inner field equations and the matching conditions, it would be 
easier to start with the simplest possible case which can be handled 
analytically. Our results show that the $q-$metric is the best candidate for 
this task. We expect to explore this problem in future works.


\begin{acknowledgements}
This work has been partially supported by the UNAM-DGAPA-PAPIIT, 
Grant No. IN111617. 
\end{acknowledgements}


\appendix

\section{The function $\gamma$ for metrics with quadrupole}
\label{sec:gamma}

In the case of the Erez-Rosen metric, the function $\gamma$ takes the form
\begin{eqnarray}
\label{gamer}
\gamma_{_{ER}} & = & \frac{1}{2} \left(q + 1 \right)^2 
\ln{ \left( {\frac{x^2 - 1}{x^2 - y^2}} \right)} \nonumber \\
& - & \frac{3}{2} q \left(1 - y^2 \right) \left[ \frac {3}{32} q 
(x^2-1) (9 \, x^2 y^2 - x^2 - y^2 + 1) 
\ln^2 \left( {\frac {x-1}{x+1}} \right) \right. \nonumber \\ 
& + & \left. \frac{1}{8} x 
( 27 \, q x^2 y^2 - 3 \, q x^2 - 21 \, q y^2 + 5\, q + 8 ) 
\ln  \left({\frac {x-1}{x+1}} \right) \right. \nonumber \\ 
& + & \left.  \frac{1}{8} (27 \, q x^2 y^2 - 3 \, q x^2 - 12 \, q y^2 
+ 4 \, q + 16) \right] .
\end{eqnarray}

For the Gutsunayev-Manko solution, the function $\gamma$ can be expressed as

\begin{eqnarray}
\label{gamgm}
\gamma_{_{GM}} & = & \frac{1}{2} \, 
\ln{ \left( {\frac {{x}^{2}-1}{{x}^{2}-{y}^{2}}} \right)} \nonumber \\ 
& + & \frac{1}{2} \, q \frac{1-y^2}{(x^2-y^2)^4} \left( 3 \, \left( 
- 5 \, {y}^{2} + 1 \right) \left({x}^{2} - {y}^{2} \right) ^{2} \right. 
\nonumber\\
& + & \left. 8 \, {y}^{2} \left( -5 \, {y}^{2} + 3 \right) 
\left( {x}^{2}-{y}^{2} \right) +24\,{y}^{4} \left( 
- {y}^{2} + 1 \right) \right)  \nonumber\\ 
& + & \frac{1}{8} \, q^2 
\frac{\left(1 - y^2 \right)}{\left({x}^{2}-{y}^{2} \right)^{8}} 
\left(- 12 \, \left( 25 \, {y}^{4} - 14 \, {y}^{2} + 1 \right) 
\left({x}^{2} - {y}^{2} \right)^{5} \right. \nonumber \\
& + & \left. 3 \, \left(- 675 \, {y}^{6} + 697 \, {y}^{4} - 153 \, {y}^{2} 
+ 3 \right) \left({x}^{2} - {y}^{2} \right)^{4} \right. \nonumber\\ 
& + & \left. 32 \, {y}^{2} \left(- 171 \, {y}^{6} + 259 \, {y}^{4}
-105 \, {y}^{2} + 9 \right)  \left({x}^{2} - {y}^{2} \right) ^{3} 
\right.\nonumber \\ 
& + & \left. 32 \, {y}^{4} \left(- 225 \, {y}^{6} + 451 \, {y}^{4} 
- 271 \, {y}^{2} + 45 \right)  \left({x}^{2} - {y}^{2} \right)^{2} \right. 
\nonumber \\ 
& + & \left. 2304 \, {y}^{6} \left(- 2 \, {y}^{6} + 5 \, {y}^{4} - 4 \, {y}^{2} 
+ 1 \right) \left({x}^{2} - {y}^{2} \right) \right. \nonumber \\ 
& + & \left. 1152 \, {y}^{8} \left(- {y}^{6} + 3 \, {y}^{4} - 3 \, {y}^{2} 
+ 1 \right) \right). 
\end{eqnarray}

The Manko solution \cite{man90} must be complemented with the function 
\begin{eqnarray}
\label{gamm}
\gamma_{_{M}} & = &  \frac{1}{2} \, 
\ln{ \left( {\frac {{x}^{2}-1}{{x}^{2}-{y}^{2}}} \right)} \nonumber \\ 
& + & \frac{q x (2 x^4 - 5 x^2 + 5 x^2 y^2 + 3 - 3 y^2)}{(x^2 + y^2 - 1)^{5/2}}
- 2 q \nonumber \\ 
& + & \frac{3 q^2}{8 (x^2 + y^2 - 1)^5} \left(\frac{x^2 y^2 
(5 x^2 y^2 - 3 x^2 - 3 y^2 + 3)^2}{x^2 + y^2 - 1} 
- (3 x^2 y^2 - x^2 - y^2 + 1)^2 \right) 
\end{eqnarray}

The Hern\'andez-Pastora-Mart{\'{\i}}n solutions are complemented by the 
following $\gamma$ functions:
\begin{eqnarray}
\label{gamhm1}
\gamma_{_{HM1}} & = & \frac{1}{2} \left(1 + \frac{225}{24} q^2 \right) 
\ln{\left(\frac{x^2 - 1}{x^2 - y^2} \right)} \nonumber \\
& - & \frac{15}{8} q x (1 - y^2) \left(1 
- \frac{15}{32} q \left[ x^2 + 7 y^2 - 9 x^2 y^2 
+ 1 - \frac{8}{3} \frac{x^2 + 1}{x^2 - y^2} \right] \right) 
\ln{\left(\frac{x - 1}{x + 1} \right)} \nonumber \\
& + & \frac{225}{1024} q^2 (x^2 - 1) (1 - y^2) (x^2 + y^2 - 9 x^2 y^2 - 1) 
\ln^2{\left(\frac{x - 1}{x + 1} \right)} \nonumber \\
& - & \frac{15}{4} q (1 - y^2) (1 - \frac{15}{64} q 
(x^2 + 4 y^2 - 9 x^2 y^2 + 4)) \nonumber \\
& - & \frac{75}{16} q^2 x^2 \frac{1 - y^2}{x^2 - y^2} 
- \frac{5}{4} q \frac{(x^2 + y^2) (1 - y^2)}{(x^2 - y^2)^2} \nonumber \\
& - & \frac{25}{64} q^2 (2 x^6 - x^4 + 3 x^4 y^2 - 6 x^2 y^2 
+ 4 x^2 y^4 - y^4 - y^6) \frac{(1 - y^2)}{(x^2 - y^2)^4} ,
\end{eqnarray}

\begin{eqnarray}
\label{gamhm2}
\gamma_{_{HM2}} & = & \frac{1}{32768} \left[\left(
\frac{15}{256} q {(1 - y^2)} 
\ln{\left[\frac{x - 1}{x + 1} \right]} \times 
\right. \right. \nonumber \\
& \times & \left. \left. \left(\frac{4 x}{(x^2 - y^2)^3}(512 (8 (5 {A}_1 q 
- 64 (x^2 - y^2)) (x^2 - y^2)^2 + 5 {A}_2 q^2) - 525 {A}_3 q^3) 
\right. \right. \right. \nonumber \\
& - & \left. \left. \left. 15 q ({A}_4 q^2 + {A}_5 q + 73728 x^2 y^2 
- 8192 x^2 - 8192 y^2 
+ 8192) (x^2 - 1) \ln{\left[\frac{x - 1}{x + 1} \right]} \right) 
\right. \right. \nonumber \\
& + & \left. \left. 32 (223875 q^4 - 268800 q^3 + 512) 
\ln{\left[\frac{x^2 - 1}{x^2 - y^2} \right]} \right) \right. \nonumber \\
& + & \left. \frac{5}{64} q (1 - y^2) \! \! \left(
\frac{512}{(x^2 - y^2)^4} \left(32 ({A}_6 q 
- 32 (3 x^4 - 6 x^2 y^2 + x^2 + 3 y^4 + y^2) (x^2 - y^2)^2) 
\right. \right. \right. \nonumber \\
& + & \left. \left. \left. \frac{5 {A}_7 q^2}{(x^2 - y^2)^2} \right)
- \frac{5}{(x^2 - y^2)^8} ({A}_8 + {A}_9 x^6 y^2) q^3 \right)
\right] ,
\end{eqnarray}
where
\begin{eqnarray*}
{A}_1 & = & 1731 x^4 y^2 - 351 x^4 + 1029 y^4 - 321 y^2 - 16 
- x^2 (1731 y^4 + 678 y^2 - 305) \nonumber \\
{A}_2 & = & 21 x^6 (5 x^2 (1785 x^2 y^4 - 1020 x^2 y^2 + 51 x^2 - 5355 y^6 
+ 850 y^4 + 1163 y^2 - 82) \nonumber\\ 
& + & 26775 y^8 + 17850 y^6 - 15490 y^4 - 450 y^2 + 723)\nonumber \\
& - & y^2 (55335 y^8 - 24570 y^6 + 11505 y^4 - 3598 y^2 - 240) \nonumber \\
& - & 3 x^4 (62475 y^{10} + 196350 y^8 - 69160 y^6 - 23100 y^4 + 14037 y^2 
- 1146)\nonumber  \\
& + & x^2 (232050 y^{10} + 45675 y^8 - 75810 y^6 + 38113 y^4 - 7036 y^2 + 80) \\
{A}_3 & = & 111517875 x^{12} y^6 - 116069625 x^{12} y^4 + 28676025 x^{12} y^2 
- 819315 x^{12} \nonumber \\
& - & 334553625 x^{10} y^8 + 157794000 x^{10} y^6 
+ 110794950 x^{10} y^4 - 45624600 x^{10} y^2 \nonumber \\
& + & 1461915 x^{10} + 334553625 x^8 y^{10} + 223035750 x^8 y^8 
- 413254275 x^8 y^6 \nonumber \\ 
& + & 48671595 x^8 y^4 
+ 18111870 x^8 y^2 - 691173 x^8 - 111517875 x^6 y^{12} \nonumber \\
& - & 455175000 x^6 y^{10} + 288232875 x^6 y^8 + 125092800 x^6 y^6 
- 52174485 x^6 y^4 \nonumber \\
& - & 567000 x^6 y^2 + 144909 x^6 - 91186725 x^2 y^{12} + 60618600 x^2 y^{10} 
+ 10071750 x^2 y^8 \nonumber \\ 
& - & 6777776 x^2 y^6 + 331047 x^2 y^4 + 107520 x^2 y^2 
+ 9600 x^2 + 10833165 y^{12} \nonumber \\
& - & 10736775 y^{10} + 2425095 y^8 - 146829 y^6 - 44160 y^4 + 28800 y^2 
\nonumber\\ 
& + & x^4 (190414875 y^{12} + 76737150 y^{10} 
- 198772245 y^8 + 33048125 y^6  \\
& + & 5959014 y^4 - 498087 y^2 - 63360) \\
{A}_4 & = & 11025 (354025 x^6 y^6 - 368475 x^6 y^4 + 91035 x^6 y^2 - 2601 x^6 
- 368475 x^4 y^6 \\
& + & 379185 x^4 y^4 - 91953 x^4 y^2 + 2907 x^4 + 91035 x^2 y^6 
- 91953 x^2 y^4 + 22257 x^2 y^2 \\
& - & 603 x^2 - 2601 y^6 + 2907 y^4 - 603 y^2 + 297) \\
{A}_5 & = & - 53760 (595 x^4 y^4 - 340 x^4 y^2 + 17 x^4 - 340 x^2 y^4 
+ 212 x^2 y^2 - 16 x^2 + 17 y^4 \\ 
& - & 16 y^2 - 1) \\
{A}_6 & = & 13185 x^{10} y^2 - 2655 x^{10} - 52740 x^8 y^4 + 7140 x^8 y^2 
+ 1440 x^8 + 79110 x^6 y^6 \\ 
& - & 2010 x^6 y^4 - 5880 x^6 y^2 + 112 x^6 - 52740 x^4 y^8 
- 10260 x^4 y^6 + 9000 x^4 y^4 \\ 
& - & 312 x^4 y^2 + 40 x^4 - 3480 y^{10} + 1560 y^8 + 152 y^6 + 40 y^4 \\
& + & x^2 y^2 (13185 y^8 + 11265 y^6 - 6120 y^4 - 272 y^2 + 240) 
\end{eqnarray*}

\begin{eqnarray*}
{A}_7 & = & 562275 x^{16} y^4 - 321300 x^{16} y^2 + 16065 x^{16} 
- 3373650 x^{14} y^6 + 1419075 x^{14} y^4 \\
& + & 211050 x^{14} y^2 - 20475 x^{14} + 8434125 x^{12} y^8 
- 1767150 x^{12} y^6 - 1439445 x^{12} y^4 \\
& + & 90930 x^{12} y^2 + 78708 x^{12} - 11245500 x^{10} y^{10} - 1204875 x^{10} y^8 
+ 3412080 x^{10} y^6 \\
& - & 144165 x^{10} y^4 - 451620 x^{10} y^2 + 1056 x^{10} 
+ 57120 y^{16} - 3360 y^{14} + 56640 y^{12} \\
& - & 1088 y^{10} - 608 y^8 - 160 y^6 
- x^2 y^4 (508725 y^{12} + 142380 y^{10} - 28245 y^8 \\
& + & 361428 y^6 - 3648 y^4 + 832 y^2 + 2400) 
+ 3 x^8 (2811375 y^{12} + 1785000 y^{10} \\ 
& - & 1326675 y^8 + 24500 y^6 + 358040 y^4 - 96 y^2 - 96) 
+ x^4 y^2 (562275 y^{14} + 2731050 y^{12} \\
& - & 436275 y^{10} - 70350 y^8 + 959940 y^6 - 2208 y^4 + 10880 y^2 - 2400) \\ 
& - & x^6 (3373650 y^{14} + 5703075 y^{12} - 2301810 y^{10} - 45675 y^8 
+ 1356360 y^6 + 6240 y^4 \\ 
& - & 1088 y^2 + 160)
\end{eqnarray*}

\begin{eqnarray*}
{A}_8 & = & 35128130625 x^{22} y^6 - 36561931875 x^{22} y^4 
+ 9032947875 x^{22} y^2 - 258084225 x^{22} \\ 
& - & 281025045000 x^{20} y^8 + 244224146250 x^{20} y^6 
- 22451640750 x^{20} y^4 \\ 
& - & 10070345250 x^{20} y^2 + 374475150 x^{20} 
+ 983587657500 x^{18} y^{10} - 637563622500 x^{18} y^8 \\ 
& - & 129817114875 x^{18} y^6 + 73875625425 x^{18} y^4 + 848869875 x^{18} y^2 
- 115835265 x^{18} \\ 
& - & 1967175315000 x^{16} y^{12} + 695871540000 x^{16} y^{10} 
+ 762842241000 x^{16} y^8 \\ 
& - & 198318141000 x^{16} y^6 - 19428973200 x^{16} y^4 
+ 718389000 x^{16} y^2 + 28304640 x^{16} \\ 
& + & 2458969143750 x^{14} y^{14} + 143858058750 x^{14} y^{12} 
- 1715997701250 x^{14} y^{10} \\ 
& + & 220631836950 x^{14} y^8 + 79818927300 x^{14} y^6 
- 1466001180 x^{14} y^4 - 266353920 x^{14} y^2 \\
& - & 29516544 x^{14} - 1967175315000 x^{12} y^{16} - 1331523427500 x^{12} y^{14} \\
& + & 2098661449500 x^{12} y^{12} + 36939968100 x^{12} y^{10} 
- 165043084500 x^{12} y^8 + 222283320 x^{12} y^6 \\
& + & 974776320 x^{12} y^4 + 163676160 x^{12} y^2 + 410112 x^{12} 
+ 983587657500 x^{10} y^{18} \\ 
& + & 1679459197500 x^{10} y^{16} 
- 1433634441750 x^{10} y^{14} - 400459377150 x^{10} y^{12} \\
& + & 200621102850 x^{10} y^{10} + 3991429050 x^{10} y^8 - 1901034240 x^{10} y^6 
- 348526848 x^{10} y^4 \\ 
& + & 4475904 x^{10} y^2 - 261120 x^{10} 
- 281025045000 x^8 y^{20} - 1059101190000 x^8 y^{18} \\
& + & 440141373000 x^8 y^{16} 
+ 499964434200 x^8 y^{14} - 146690964000 x^8 y^{12} - 7673340360 x^8 y^{10} \\
& + & 2224588800 x^8 y^8 
+ 299837184 x^8 y^6 - 7544064 x^8 y^4 - 1839360 x^8 y^2 - 57600 x^8 \\
& - & 48271308750 x^4 y^{22} 
- 76235118750 x^4 y^{20} + 92851469550 x^4 y^{18} - 8613319050 x^4 y^{16} \\
& - & 3798768120 x^4 y^{14} 
+ 780702720 x^4 y^{12} - 257951232 x^4 y^{10} + 5118976 x^4 y^8 \\
& + & 12577280 x^4 y^6 - 4032000 x^4 y^4 + 15755882625 x^2 y^{22} 
- 9452779875 x^2 y^{20} \\ 
& - & 2139220125 x^2 y^{18} + 1125993855 x^2 y^{16} 
- 226571520 x^2 y^{14} + 169069824 x^2 y^{12} \\ 
& + & 2873344 x^2 y^{10} - 3266560 x^2 y^8 - 1612800 x^2 y^6 - 815673600 y^{22} 
+ 719712000 y^{20} \\ 
& - & 147974400 y^{18} + 31799040 y^{16} 
- 37271808 y^{14} - 284928 y^{12} - 344320 y^{10} - 57600 y^8 \\
{A}_9 & = & 35128130625 y^{20} + 349608538125 y^{18} + 51702123375 y^{16} 
- 304888933125 y^{14} \\ 
& + & 59532473700 y^{12} + 7143824100 y^{10} - 1646211840 y^8 + 48056064 y^6 
- 27167744 y^4 \\ 
& + & 15252480 y^2 - 1612800
\end{eqnarray*}

\section{The Kretschmann scalar of the Erez-Rosen spacetime}
\label{sec:singer}

For the  general metric in prolate spheroidal coordinates (\ref{lelxy}), 
the Kretschmann scalar can be represented as 
\begin{equation}
K = \frac{48}{\sigma^4} \, \frac{ \mathcal{P}(x,y;\partial \psi,
\partial^2 \psi,\partial \gamma,\partial^2 \gamma) }{ (x^2-1) (x^2-y^2)^3 
(1-y^2) } \, {\rm e}^{4(\psi-\gamma)},
\end{equation}
where $\mathcal{P}$ is a polynomial function in each of its arguments and 
$\partial^m \psi$ and $\partial^m \gamma$ represent the $m$-th derivatives 
of $\psi$ and $\gamma$ with respect to $x$ and $y$. The field equations for 
$\gamma$ can be used to express all the first and second derivatives of 
$\gamma$ in terms of those of $\psi$. Using these relations, the 
Kretschmann scalar becomes
\begin{equation}
K = \frac{48}{\sigma^4} \, \frac{ P(x,y;\partial \psi,\partial^2 \psi) }{ 
(x^2-y^2)^4 } \, {\rm e}^{4(\psi-\gamma)},
\end{equation}
where $P$ is also a polynomial function in each of its arguments. The 
explicit forms of the polynomial functions $\mathcal{P}$ and $P$ 
is not given here  since they involve very long expressions which do not 
provide any insight for the present analysis. We see that, when written in 
this form, the Kretschmann scalar shows only one singularity when 
$x^2-y^2 =0$. Nevertheless, the behavior of the polynomial function and the 
exponential factor could cancel this divergence or introduce new ones. 

For metrics  whose function $\psi$ depends on both $x$ and $y$, the analysis 
of the Kretschmann scalar reduces to the analysis of 
the polynomial $P(x,y,\partial \psi,\partial^2 \psi)$ and the exponential 
${\rm e}^{4(\psi-\gamma)}$ and their relationship with the factor 
$(x^2-y^2)^{-4}$. The Schwarzschild metric serves as a guide mark to see what 
kind of behavior to expect from the polynomial and exponential factors in 
$K$. In this case, the metric functions are given by Eq.(\ref{schw}) that
correspond to ${\rm e}^{4(\psi-\gamma)} = (x+1)^{-4} (x^2-y^2)^2$ and 
$P(x,y;\partial \psi,\partial^2 \psi) = (x+1)^{-2} (x^2-y^2)^2$, leading 
to the scalar (\ref{kschw}). This means that both the exponential and the 
polynomial factors contribute to generate the divergence 
at $x=-1$, and also to cancel out the original factor that diverges at 
$x=\pm y$.

In the case of other static axisymmetric metrics, the exponential and the 
polynomial factors produce new divergent factors or factors that cancel the 
original divergent factor. Let us consider in detail the case of the 
Erez-Rosen metric. The exponential and polynomial factors are
\begin{eqnarray}
{\rm e}^{4(\psi-\gamma)} & = & \frac{\exp \left[ \displaystyle\sum_{n=0}^2 
\pi_n(x,y;q) \left( \ln\frac{x-1}{x+1} \right)^n\right]}
{(x+1)^{2(q^2+2q+2)} (x-1)^{2q(q+2)} (x^2-y^2)^{-2(q+1)^2}},\\
P(x,y;\partial \psi,\partial^2 \psi) & = & \frac{x^2-y^2}{(x^2-1)^2} \, 
\displaystyle\sum_{n=0}^6 p_n(x,y;q)\left( \ln\frac{x-1}{x+1} \right)^n \ ,
\end{eqnarray}
where $\pi_n$ and $p_n$ are polynomial functions of each of its arguments 
with the following properties
\begin{align*}
\pi_n(x,y;0) & = 0 ,& \forall n,\\
p_n(x,y;q) & = q^n \tilde{p}_n(x,y;q) ,& \forall n,\\
\pi_2(\pm 1,y;q) & = 0,\\
p_n(\pm 1,y;q) & = 0, & n>0,\\
p_n(x=\pm y,y;q) & = 0 , & n \geq 4,
\end{align*}
so that the Kretschmann can be written as
\begin{equation}
K_{_\text{ER}} = \frac{48}{\sigma^4} \frac{\left[ \displaystyle\sum_{n=0}^6 
p_n(x,y;q) \left( \ln\frac{x-1}{x+1} \right)^n \right] 
 \exp\left[ \displaystyle\sum_{n=0}^2 \pi_n(x,y;q) 
\left( \ln\frac{x-1}{x+1} \right)^n   \right] }
{(x+1)^{2(q^2+2q+3)} (x-1)^{2(q+1)^2} (x^2-y^2)^{-2q^2-4q+1}} \ . 
\end{equation}

For the particular cases $q=0$, $q=-1$ and $q=-2$, 
 the factor $\displaystyle\sum_{n=0}^6 p_n(x,y;q) $$ 
\left( \ln\frac{x-1}{x+1} \right)^n$ vanishes for $x=\pm 1$, or $x=\pm y$. 
Indeed, for $q=0$, we have that $\displaystyle\sum_{n=0}^6 p_n(x,y;0)$$ 
\left( \ln\frac{x-1}{x+1} \right)^n $$= p_0(x,y;0) = (x-1)^2 (x^2-y^2)$, and 
using also that $\pi_n(x,y;0)=0$, we obtain
$K_\text{ER} \left.\right|_{q=0} = \frac{48}{\sigma^4} \frac{1}{(x+1)^6}$, 
which corresponds to the Kretschmann scalar of the Schwarzschild metric, as 
expected. For $q=-1$, it turns out that all the polynomials $p_n(x,y;-1)$ 
have roots at $x= \pm y$ with multiplicity two for $0 \leq n \leq 3$ 
and multiplicity three for $n \geq 4$. Therefore, the term 
$(x^2-y^2)^2$ is canceled with the one in the denominator and the 
Kretschmann scalar turns out to be
\begin{equation}
K_\text{ER} \left.\right|_{q=-1} = \frac{48}{\sigma^4} \frac{ \left[ 
\displaystyle\sum_{n=0}^6 \bar{p}_n(x,y) \left( \ln\frac{x-1}{x+1} 
\right)^n\right]  
\exp \left[  \displaystyle\sum_{n=0}^2 \pi_n(x,y;-1) 
\left( \ln\frac{x-1}{x+1} \right)^n \right]}
{(x+1)^4 (x^2-y^2)} .
\end{equation}
For $q=-2$, the polynomials $p_n(x,y;-2)$ have roots at $x= \pm y$, 
but now with multiplicity one for $0 \leq n \leq 4$, multiplicity two for 
$n=5$ and multiplicity three for $n=6$. Then, in a similar way, the term 
$x^2-y^2$ is canceled so that
\begin{equation}
K_{_\text{ER}} \left.\right|_{q=-2} = \frac{48}{\sigma^4} \frac{ \left[ 
\displaystyle\sum_{n=0}^6 \bar{\bar{p}}_n(x,y) \left( \ln\frac{x-1}{x+1} 
\right)^n\right]
\exp\left[ \displaystyle\sum_{n=0}^2 \pi_n(x,y;-2) \left( 
\ln\frac{x-1}{x+1} \right)^n \right] 
 }{(x+1)^6 (x-1)^2}\  .
\end{equation}

For values of $q$, different from the ones analyzed above, the Kretschmann 
scalar for the Erez-Rosen metric presents divergences at
 $x=\pm 1$ and $x=\pm y$, if the exponents of the corresponding factors 
are negative and if the exponential factor does not vanish for those values, 
in which case one would have to calculate the limit explicitly. The 
exponent of the factors $x-1$ has a negative definite sign and so there is 
a curvature singularity at $x=-1$. On the contrary, the exponent of the 
factor $x^2-y^2$ vanishes for $q=-1 \pm \sqrt{3/2}$, indicating that for certain 
ranges of values of $q$ there could be curvature singularities at 
$x= \pm y$. 

The case $x=1$ deserves a detailed analysis for which  we have that 
\begin{eqnarray}
\left. \displaystyle\sum_{n=0}^6 p_n(x,y;q) \left( \ln\frac{x-1}{x+1} \right)^n \right|_{x=1} 
& = & p_0(1,y;q),\\
\left. \displaystyle\sum_{n=0}^2 \pi_n(x,y;q) \left( \ln\frac{x-1}{x+1} \right)^n \right|_{x=1}
& = & \pi_0(1,y;q) + \left. \pi_1(x,y;q) \ln\frac{x-1}{x+1} \right|_{x=1},
\end{eqnarray}
so that taking the limit, we obtain 
\begin{equation}
\lim_{x \to 1} K_{_\text{ER}} = \frac{48}{\sigma^4} \frac{p_0(1,y;q) \, 
{\rm e}^{\pi_0(1,y;q)}}{2^{2(q^2+2q+3)+\pi_1(1,y;q)} (1-y^2)^{-2q^2-4q+1}} \lim_{x \to 1} 
(x-1)^{\pi_1(1,y;q)-2(q+1)^2}. 
\end{equation}
This limit vanishes for $\pi_1(1,y;q)-2(q+1)^2 \geq 0$ and diverges for 
$\pi_1(1,y;q)-2(q+1)^2 < 0$. Explicitly $\pi_1$ is given by 
$\pi_1(1,y;q) = \frac{3}{2} q \left\{ q\left[1+3y^2
\left(\frac{2}{3}-y^2\right) \right] + 2\left( \frac{5}{3}-y^2 \right) 
\right\}$ so that the exponent can be written as 
$\pi_1(1,y;q)-2(q+1)^2 = -\frac{1}{2}([q(3y^2-1)+1]^2 + 3)$ 
that is strictly negative, meaning the Kretschmann scalar diverges for $x=1$.
The {\it sign} of the divergence will be determined by $p_0(1,y;q)$, which explicitly is given by $p_0(1,y;q) = \frac{1}{192} q^2 (1-y^2) \, (3y^2-1)^2 \, (q[3y^2-1]+2)^2 \, ([q(3y^2-1)+1]^2 + 3)$, that given the range of $y$, $y \in (-1,1)$, is positive definite, hence
\begin{equation}
\lim_{x \to 1} K_\text{ER} = + \infty.
\end{equation}

We thus conclude that the Erez-Rosen spacetime has a singularity at $x=- 1$, 
independently of the value of $q$ and the coordinate $y$. 
Then, for certain values of $q$, there is a second singularity at $x=\pm y$ 
and, finally, at $x=1$ there could be a third singularity, depending on the 
value of $q$. All these singularities are naked in the sense that they are not 
covered by an exterior horizon. This means that for particular values of $q$, 
it is possible to observe the singularity located at the origin of coordinates 
$x=-1$.

\section{ Explicit expressions for the multipole moments }
\label{sec:mm}

The multipole moments $M_n$ for a given solution can be obtained from the derivatives of the Ernst potential evaluated at the axis of symmetry, $m_n$, plus an additional term $d_n$ which is different for each $n$ and can be expressed in terms of $m_n$. The additional terms for $n=1,...,10$ are expressed as:

\bea
d_0 & = & d_1 = d_2 = d_3 = 0 \ , \nonumber \\
d_ 4 & = &  \frac{1}{7}  m_0(m_1^2 - m_2 m_0) \ , \nonumber \\
d_5 & = & \frac{1}{3} m_0 (m_2 m_1 - m_3 m_0) 
+ \frac{1}{21} m_1 (m_1^2-m_2 m_0) \ , \nonumber \\
d_6 & = & \frac{2}{11} m_0 (m_3 m_1 - 3 m_4 m_0) 
+ \frac{1}{33} m_0^3 (m_2 m_0 + m_1^2) 
+ \frac{1}{77} m_2 (11 m_1^2 + 17 m_0 m_2) \ , \\
d_7 & = & \frac{1}{39} m_0 (18 m_3 m_2 - 33 m_5 m_0 - 4 m_2 m_1 m_0^2) 
\nonumber \\
& + & \frac{1}{429} (12 m_4 m_1 m_0 + 51 m_3 m_1^2 + 45 m_3 m_0^4 
+ 69 m_2^2 m_1 - m_1^3 m_0^2) \ , \nonumber \\
d_8 & = & - m_6 m_0^2 
+ \frac{1}{39} (3 m_4 m_1^2 + m_0 (9 m_4 m_0^3 - 6 m_5 m_1 + 11 m_3^2)) 
\nonumber \\
& + & \frac{1}{429}(m_0 (180 m_4 m_2 - 36 m_3 m_1 m_0^2 - 3 m_2 m_0^5 
+ 3 m_1^2 m_0^4) + m_2 (138 m_3 m_1 + 23 m_2^2)) \nonumber \\ 
& + & \frac{1}{3003} m_0 (31 m_1^4 - 382 m_2^2 m_0^2 - 90 m_2 m_1^2 m_0) 
\ , \nonumber \\
d_9 & = & \frac{1}{17} m_0 (7 m_5 m_0^3 - 21 m_7 m_0 - 6 m_6 m_1) 
+ \frac{1}{11} m_2 \left(\frac{23}{13} m_3 m_2 
+ \frac{6}{17} m_1 m_0^5 \right) \nonumber \\ 
& + & \frac{1}{221} (76 m_5 m_2 m_0 + 5 m_5 m_1^2 + 64 m_4 m_2 m_1 
+ 4 m_4 m_1 m_0^3 - 80 m_3 m_2 m_0^3 - 7 m_3 m_0^6) \nonumber \\ 
& + & \frac{1}{2431} (1432 m_4 m_3 m_0 + 443 m_3^2 m_1 - 126 m_3 m_1^2 m_0^2 
- m_1^3 m_0^4) \nonumber \\
& + & \frac{1}{17017} m_1 (41 m_1^4 
- 1002 m_2^2 m_0^2 + 688 m_2 m_1^2 m_0) \ , \nonumber \\
d_{10} & = & \frac{1}{323} (210 m_6 m_0^4 - 476 m_8 m_0^2 
- 182 m_7 m_1 m_0 + 80 m_6 m_2 m_0 - 13 m_6 m_1^2 \nonumber \\
& + & 70 m_5 m_1 m_0^3 - 28 m_4 m_0^6) 
+ \frac{1}{4199} (2406 m_5 m_3 m_0 + 982 m_5 m_2 m_1 + 699 m_3^2 m_2 
\nonumber \\
& + & 126 m_3 m_1 m_0^5 + 7 m_2 m_0^8 - 7 m_1^2 m_0^7)
+ \frac{1}{13} m_1^4 \left(\frac{205}{1309} m_2 
- \frac{50}{969} m_0^3 \right) \nonumber \\
& + & \frac{1}{46189} (15319 m_4^2 m_0 + 17198 m_4 m_3 m_1 
+ 7039 m_4 m_2^2 - 19406 m_4 m_2 m_0^3 \nonumber \\ 
& - & 1439 m_4 m_1^2 m_0^2 
- 10942 m_3 m_2 m_1 m_0^2) 
+ \frac{1}{138567} m_0 (3700 m_3 m_1^3 - 39317 m_3^2 m_0^2 \nonumber \\
& + & 7589 m_2^2 m_0^4 + 815 m_2 m_1^2 m_0^3) 
+ \frac{1}{969969} m_2^2 m_0 (66930 m_1^2 - 2609 m_2 m_0) 
 \ . \nonumber 
\eea


\end{document}